\documentclass[twocolumn,pre,amsmath]{revtex4}
\usepackage{graphicx}
\usepackage{dcolumn}
\usepackage{amsmath}
\usepackage{amssymb}
\usepackage{amsfonts}
\usepackage{amscd}
\begin{document}
\title{The Diffusion of the Magnetization Profile in the $XX$-model}
\author{Yoshiko Ogata}
\address{Department of Physics, Graduate School of Science, 
 University of Tokyo,
  Hongo,7-3-1,Bunkyo-ku, Tokyo 113-0033, Japan}
\email{ogata@monet.phys.s.u-tokyo.ac.jp}
\date{\today}
\begin{abstract}
By  the $C^*$-algebraic method, we investigate the 
magnetization profile in the intermediate time of diffusion.
We observe a transition from monotone profile to
non-monotone profile.
This transition is purely thermal. 
\end{abstract}
\maketitle
\section{Introduction}
The anomalous properties of 
the state with current in
one-dimensional integrable
system 
has attracted considerable interests.
Especially, the heat conduction in one-dimensional
systems is a long -standing problem.
It has been expected that the existence of conserved quantities implies
anomalous conductivity of the heat current \cite{zotos1997}.
A large number of numerical investigations have been done
\cite{savin02}

It is natural to consider the state with current
as the non-equilibrium steady state (NESS),
i.e., the state which is asymptotically realized 
from the inhomogeneous initial state \cite{rue:natu},\cite{pil:non}.
In other words, the NESS is the state at the convergent point.
There reported many works on the NESS itself.
However, the relaxation process to the NESS is 
not much investigated.
After a large but finite time, what kind of 
profile does the observable quantity show? 
In the previous paper \cite{yosi1}, stating from the
inhomogeneous initial state (the temperatures of the left and the right 
are different), we obtained the homogeneous NESS
for the transverse $XX$-model.
Investigating the profiles in the intermediate time,
we would obtain the diffusion of the temperature profile.
Furthermore, in the integrable system, 
the existence of the conserved quantities
may have a significant
influence on the relaxation process.

In this paper, we investigate the profile at the large intermediate time
using the transverse $XX$-model.
The chain is initially divided to the left and the right,
and kept at different temperatures.
The problem of intermediate profile with inhomogeneous initial condition
in the $XX$-model 
was first studied by Antal.et.al. in \cite{ant2}.
They considered the magnetization profile
in zero temperature with the reversed external field,
the situation which makes the calculation simple.
It was shown that the magnetization profile shows the scaling property,
$m(x,t)\approx\Phi(x/t)$.
They calculated the explicit form of the scaling function
$\Phi$, which has the flat part
around the origin.
We investigate this problem with the aid of the $C^*$-algebraic argument.
Modifying the argument of Ho and Araki \cite{HA00},
we develop a method which is applicable to 
the present situation.
The $C^*$-algebraic argument makes the calculation much simpler, and
enables us to consider the more general situation,
including the finite temperature case.

We obtain the scaling limit $m(x,t)\approx\Phi(x/t)$.
For the situation of Antal.et.al. \cite{ant2},
we reproduce the same result.
We further consider the finite temperature case.
In the finite temperature case, we obtain a remarkable dependence 
of the magnetization profile on the strength of the external field
and the temperature.
When the external field is large, or the difference of the temperature is small,
the profile varies monotonically.
On the other hand, when the external field is small and the difference of the temperature is large,
the profile is not monotone and has two extremum points.
This feature is absent at the zero-temperature.
That is, this phenomenon is due to purely thermal effect.
This can be explained by the velocity distribution.
In section \ref{model},we represent the model and the initial condition.
In section \ref{scaling}, the scaling property of correlation function is derived.
In section \ref{profile}, we investigate the interesting property which 
the profile reveals.

\section{The model}\label{model}
The Hamiltonian we shall consider has the form,
\begin{eqnarray}
{\cal H}&=&\frac{1}{4}\sum_{n=-\infty}^{\infty}
\left(
{\sigma}_{n}^{x}{\sigma}_{n+1}^{x}
+{\sigma}_{n}^{y}{\sigma}_{n+1}^{y}\right)
+\frac{\gamma}{2}\sum_{n=-\infty}^{\infty}{\sigma}_{n}^{z} ,
\label{hamil}
\end{eqnarray}
where $\sigma_{n}^{\alpha}\, (\alpha = x,y,z)$ is 
the $\alpha$-component of the Pauli matrix at the site $n$.
This Hamiltonian is called the transverse $XX$-model.
The Hamiltonian is
written by the fermion operators using the Jordan-Wigner transformation
\cite{lieb};
\begin{eqnarray}
{\cal H}&=
-\frac12 \sum_{n=-\infty}^{\infty}
\left[a_{n+1}^{\dag}a_{n}+
a_{n}^{\dag}a_{n+1}\right]\nonumber\\
&+\frac{\gamma}{2}
\sum_{n=-\infty}^{\infty}\left(2a_{n}^{\dag}a_{n}-1 \right) ,
\label{ahamil}
\end{eqnarray}
where $a_{n}$ and $a_{n}^{\dag}$ are the 
fermionic annihilation and creation 
operators on the $n$th site.

Owing to the bilinear form of the Hamiltonian,
the dynamics of the many body system can be written by
the dynamics of the single particle.
So, we only have to consider the dynamics of
the linear combination of $a_l^{\dag}$;
\begin{align}
a^{\dag}\left( f \right)
\equiv\sum_{l=-\infty}^{\infty}
f(l)a_l^{\dag},\;\;\;\;\;
\sum_{l=-\infty}^{\infty}\vert f(l)\vert^2<\infty.
\label{al}
\end{align}
Here, $l$ denotes the $l$-th site.
The summable sequence $\{ f(l) \}$ can be interpreted as
the wave function of the single particle which is
living on the lattice.
They construct the one particle Hilbert space,
with the inner product
\begin{align*}
\langle f\vert g\rangle =
\sum_{l=-\infty}^{\infty}{\overline{f(l)}}g(l).
\end{align*}
As usual, we define the norm of $f$ as
\begin{align*}
\Vert f\Vert=\left(\sum_{l=-\infty}^{\infty}\vert f(l)\vert^2\right)^{\frac 12}
\end{align*}

In the one particle Hilbert space,
the Fourier transformation is defined as
\begin{align*}
{\hat f}(k)\equiv
\sum_{n=-\infty}^{\infty}
f(n){\rm e}^{-ink},\quad
f(n)=\frac{1}{2\pi}
\int_{-\pi}^{\pi}
{\hat f}(k){\rm e}^{-ink}.
\end{align*}

By the Hamiltonian (\ref{ahamil}), $a^{\dag}(f)$ evolves as
\begin{align}
a^{\dag}(f)\to a^{\dag}({\rm e}^{ith}f),
\label{1d}
\end{align}
where ${\rm e}^{ith}$ is the
dynamics of the single particle, 
represented in the Fourier representation as
\begin{align*}
{\widehat{{\rm e}^{ith}f}}(k)
={\rm e}^{-it(\cos k-\gamma)}{\hat f}(k).
\end{align*}
Hence, in the coordinate representation, we have
\begin{align}
({\rm e}^{ith} f)(n)
=\frac{1}{2\pi}
\int_{-\pi}^{\pi}dk{\rm e}^{-it(\cos k-\gamma)}
{\rm e}^{ink}{\hat f}(k).
\label{ho}
\end{align}

From now on, we use the following notations. 
Usually, the Heisenberg representation of the observable $A$ is
\begin{align*}
{\rm e}^{itH}A{\rm e}^{-itH}.
\end{align*}
In stead of this, we use the notation $\alpha_t(A)$;
\begin{align*}
{\rm e}^{itH}A{\rm e}^{-itH}\leftrightarrow
\alpha_t(A).
\end{align*}
In this notation, the dynamics of the single particle
\ref{1d} is written as
\begin{align*}
\alpha_t(a^{\dag}(f))=a^{\dag}({\rm e}^{ith}f).
\end{align*}

We also use the unusual notation about the states.
Usually, the state is given by a density matrix $\rho$,
and the expectation value of $A$ is given by
\begin{align*}
{\rm Tr}\rho A.
\end{align*}
In stead of this, we use $\omega$;
\begin{align*}
{\rm Tr}\rho A
\leftrightarrow
\omega(A).
\end{align*}
These notations are introduced because the usual notations are
mathematically ill-defined in infinite system.
However, there is no inconvenience in interpreting them
in the usual sense. 

The initial state $\omega_{0}$ we consider is inhomogeneous.
To define it, we divide the chain to the left and the right.
The left $\left(n \le 0 \right)$ 
side is in the equilibrium under magnetic field 
$\gamma_-$ with inverse temperature $\beta_-$.
We denote the state by $\omega_-$.
On the other hand, the right $\left(n \ge 1 \right)$ 
side is in the equilibrium under magnetic field 
$\gamma_+$ with inverse temperature $\beta_+$.
We denote the state by $\omega_+$.
The expectation value of 
$a^{\dag}(g)a(f)$ in $\omega_-,\omega_+$
are expressed as
\begin{align}
\omega_-(a^{\dag}(g)a(f))
&=\frac{1}{\pi} 
\int_{-\pi}^{\pi} dk
\rho_-(k){\tilde g}_-(k){\overline {\tilde f}}_-(k),
\nonumber\\
\omega_+(a^{\dag}(g)a(f))
&=\frac{1}{\pi} 
\int_{-\pi}^{\pi} dk
\rho_+(k){\tilde g}_+(k){\overline {\tilde f}}_+(k),
\label{initial}
\end{align}
where $\rho_{\pm}$ is determined by the valuables $\beta_{\pm}$,
and $\gamma_{\pm}$ as
\begin{align*}
\rho_r(k)=
\frac{1}{1+e^{-\beta_{r}
\left( \cos\left(k \right)-\gamma_r \right)}}.
\end{align*}
We used the Fourier-sine transform;
\begin{align*}
{\tilde f}_-(k)&\equiv
-i\sum_{n=-\infty}^{0}f(n)\sin ((n-1)k)\quad k\in [0,\pi],\\
f(n)&=\frac{2i}{\pi}
\int_0^{\pi}dk{\tilde f}_-(k)\sin ((n-1)k)\quad n\le 0,\\
{\tilde f}_+(k)&\equiv
-i\sum_{n=1}^{\infty}f(n)\sin (nk)\quad k\in [0,\pi],\\
f(n)&=\frac{2i}{\pi}
\int_0^{\pi}dk{\tilde f}_+(k)\sin (nk)\quad n\ge1.
\end{align*}
Because of the quadratic form of the equilibrium states
$\omega_-$ and $\omega_+$,
the expectation values are evaluated by use of 
the Wick product with the two-point function.
The initial state has then the following product form; 
\begin{eqnarray}
\omega_{0}\left(A_-\otimes A_+ \right)
=\omega_-\left(A_- \right)
\omega_+\left(A_+ \right) ,
\end{eqnarray}
where $A_-$ and $A_+$ are arbitrary operators of 
the left part $\left(n \le 0 \right)$ 
and the right part $\left(n \ge 1 \right)$ of the chain, respectively.
\section{The Scaling Property}\label{scaling}

Now, we investigate the asymptotic profile of physical quantities.
Let us consider $X_n$, some physical value localized in the neighborhood of site
$n$.
The expectation value of $X_n$ at the time
$t$ is $X(n,t)\equiv\omega_0(\alpha_t(X_n))$,
with the notation of the previous section.
The scaling property means
\begin{align*}
X(n,t)\approx \Phi_X\left(\frac{n}{t}\right),
\end{align*}
i.e., for large $t$, 
the expectation value of $X$
at the site $vt$ is almost $\Phi_X(v)$. 
Fig.1 shows the situation.
Each figure is the snapshot of the profile of some $X$, at the time $t_0$, $3t_0$, $10t_0$.
Note that the leaned area has the width $t_0$, $3t_0$, $10t_0$, respectively.
It shows that the {\it wave} of the diffusion expand with the constant
velocity $1$.They can be written as
\begin{align*}
X(n,t)=\Phi_X\left(\frac{n}{t}\right),
\end{align*}
Here, $\Phi_X(v)$ is the scaling function represented in Fig.2.

To show this scaling property, we have to prove
the following convergence 
to the scaling function $\Phi_X$;
\begin{align*}
\Phi_X\left(v\right)=
\lim_{t\to\infty}X(vt,t)=
\lim_{t\to\infty}\omega_0(\alpha_t(X_{vt}))
.
\end{align*}
As $\omega_0$ is determined by the Wick product of 
the two point function, we only have to
derive the limit for $X=a^{\dag}(g)a(f)$ case;
\begin{align*}
\omega_{v}(a^{\dag}(g)a(f))\equiv\Phi_{a^{\dag}(g)a(f)}(v)=
\lim_{t\to\infty}\omega_0(\alpha_t(a^{\dag}(S_{vt}g)a(S_{vt}f)))
\end{align*}
where we used the $m$-sift operator $S_m$
on the one particle Hilbert space;
\begin{align*}
(S_mf)(n)\equiv f(n-m).
\end{align*}
For example, to derive the asymptotic profile of the magnetization
$m(v)$,
we calculate
\begin{align*}
&m\left(v\right)=
\lim_{t\to\infty}\omega_0\left(\alpha_{t}\left(\frac12\sigma_{vt}^z\right)\right)\\&=
\lim_{t\to\infty}\omega_0\left(\alpha_{t}\left(a_{vt}^{\dag}a_{vt}\right)\right)-\frac12\\ &=
\lim_{t\to\infty}\omega_0\left(\alpha_{t}
\left(a^{\dag}\left(S_{vt}\eta_0\right)a\left(S_{vt}\eta_0\right)\right)\right)-\frac12\\
&=\omega_{v}\left(a^{\dag}\left(\eta_0\right)a\left(\eta_0\right)\right)-\frac12,
\end{align*}
with $\eta_0$, a wave function defined as
\begin{align*}
\eta_0(n)=
\left\{
\begin{gathered}
1\quad n=0\\
0\quad n\neq 0.
\end{gathered}
\right.
\end{align*}
\begin{figure}[t]
\setlength{\unitlength}{0.2mm}
\begin{picture}(225,320)(0,-225)
\put(95,40){\thicklines\line(1,2){10}}
\put(-70,40){\thicklines\line(1,0){165}}
\put(105,60){\thicklines\line(1,0){165}}
\put(95,33){\thicklines\vector(-1,0){10}}
\put(105,67){\thicklines\vector(1,0){10}}
\put(300,45){$\Huge t_0$}
\put(300,10){$\Huge \Downarrow$}
\put(95,25){\line(0,-1){15}}
\put(105,25){\line(0,-1){15}}
\put(95,18){\line(1,0){10}}
\put(95,0){$t_0$}
\put(-85,35){-1}
\put(280,55){1}

\put(85,-50){\thicklines\line(3,2){30}}
\put(-70,-50){\thicklines\line(1,0){155}}
\put(115,-30){\thicklines\line(1,0){155}}
\put(85,-57){\thicklines\vector(-1,0){10}}
\put(115,-23){\thicklines\vector(1,0){10}}
\put(300,-45){$3t_0$}
\put(300,-80){$\Huge \Downarrow$}
\put(85,-75){\line(0,-1){15}}
\put(115,-75){\line(0,-1){15}}
\put(85,-82){\line(1,0){30}}
\put(95,-100){$3t_0$}
\put(-85,-65){-1}
\put(280,-35){1}

\put(50,-140){\thicklines\line(5,1){100}}
\put(-70,-140){\thicklines\line(1,0){120}}
\put(150,-120){\thicklines\line(1,0){120}}
\put(50,-147){\thicklines\vector(-1,0){10}}
\put(150,-113){\thicklines\vector(1,0){10}}
\put(300,-135){$10t_0$}
\put(50,-165){\line(0,-1){15}}
\put(150,-165){\line(0,-1){15}}
\put(50,-172){\line(1,0){100}}
\put(95,-190){$10t_0$}
\put(-85,-155){-1}
\put(280,-125){1}
\put(-70,-210){{\small
Fig.1:The picture of diffusion.
The $x$-axes is the site.
}}
\put(-70,-225){{\small
Each picture shows the
snapshot of the profile of the
}}
\put(-70,-240){{\small
local physical quantity
$X$ at the time $t_0,3t_0,10t_0$. }}
\put(-70,-255){{\small The wave of the diffusion
spread with the velocity $1$.}}
\end{picture}
\end{figure}
\begin{figure}[b]
\setlength{\unitlength}{0.2mm}
\begin{picture}(225,100)(0,0)
\put(95,40){\thicklines\line(1,2){10}}
\put(-70,40){\thicklines\line(1,0){165}}
\put(105,60){\thicklines\line(1,0){165}}
\put(95,33){\thicklines\vector(-1,0){10}}
\put(105,67){\thicklines\vector(1,0){10}}
\put(95,25){\line(0,-1){15}}
\put(95,0){$1$}
\put(105,25){\line(0,-1){15}}
\put(95,18){\line(1,0){10}}
\put(-85,35){-1}
\put(280,55){1}
\put(-10,-20){\vector(1,0){200}}
\put(100,-40){$v$}
\put(-70,-60){{\small
Fig.2
The scaling function $\Phi_X(v)$ corresponds to Fig.1.
In this}}
\put(-70,-75){{\small
 figure, the $x$-axes is not sites,
but the scaling factor $v$.
}}
\end{picture}
\end{figure}
Using the $C^*$-algebraic method, we can make the argument generic.
As an advantage, we can treat the finite temperature case.
The argument uses the result of Ho and Araki \cite{HA00}.
They calculated the following decomposition of (\ref{ho})
with $f$:analitic function;
for $\vert n\vert < t(1-\delta)$, 
with $\delta>0$, we have
\begin{align*}
({\rm e}^{ith} f)(n) =(T_tf)(n)+(Af)(n,t).
\end{align*}
$T_t$ is the operator which is defined as
\begin{align}
&(T_tf)(n)\nonumber\\
&=\left\{
\begin{gathered}
g(n,t)\;\;\;
{\rm for}\;\;\vert n\vert < t(1-\delta)\\
({\rm e}^{ith}f)(n)\;\;\;
{\rm for}\;\;\vert n\vert \ge t(1-\delta)\\
\end{gathered} 
\right.
\label{tt}
\end{align}
where
\begin{multline*}
g(n,t)\equiv
\frac12 \left(\frac{\pi}{2} t\sqrt{1-\left(n/t\right)^2} \right)^{-\frac12}\\
\times \left[{\hat f}\left(-\sin^{-1}(n/t)\right)
{\rm e}^{-in\sin^{-1}(n/t)-it\sqrt{1-(n/t)^2} +i\pi /4}\right.\\
\left.
+{\hat f}(\pm \pi+\sin^{-1}(n/t))
{\rm e}^{in(\pm \pi+ \sin^{-1}(n/t))+it\sqrt{1-(n/t)^2} -i\pi /4}
\right]\\
(+ \, {\rm for} \, n\le 0 ,\,\,- \, {\rm for}\, n>0).
\end{multline*}
This term corresponds to the contribution from the momentum $k(n,t)$
where the phase velocity
\begin{align*}
\phi(k)=-\cos k +\gamma +\frac{nk}{t}
\end{align*}
is stationary i.e., 
\begin{align}
&\phi'\left(k\right)=\sin k+\frac nt =0\nonumber\\
&\Rightarrow
k(n,t)=-\sin^{-1}(n/t),\pm\pi+\sin^{-1}(n/t)\nonumber\\
&+ \, {\rm for} \, n\le 0 ,\,\,- \, {\rm for}\, n>0
.
\label{dash}
\end{align}
$(Af)(n,t)$ decay as
\begin{align}
\vert (Af)(n,t) \vert \le \frac {C^{\delta}}{t}
\label{af}
\end{align}
with $C^{\delta}$, some constant which is independent of $n$.
They also showed that the contribution 
to ${\rm e}^{ith}f$ from $\vert n\vert > t(1-\delta)$;
\begin{align}
B_{\delta}\equiv\lim_{t\to\infty}
\sum_{\vert n\vert > t(1-\delta)}
\vert ({\rm e}^{ith}f)(n)\vert^2 \to 0
\label{bd}
\end{align}
goes to zero as $\delta\to 0$.
(\ref{af}), (\ref{bd}) ensures the following;
\begin{align}
{\rm e}^{ith}f\sim T_tf,
\label{res} 
\end{align}
for large $t$.
We can see from (\ref{tt}),
this means that at the site $n$, there is only 
the particle at momentum $k(n,t)$.

Now, let us return to the current problem.
First, we define projection operators 
on the one particle Hilbert space.
The coordinate projections $P_m^{-},P_m^{+}$
are
\begin{align*}
[{P}_m^-f](n)&\equiv
\left\{
\begin{gathered}
0\;\;\; n>m\\
f(n)\;\;\;\;n\le m
\end{gathered}
\right.\\
[{P}_m^+f](n)&\equiv
\left\{
\begin{gathered}
f(n)\;\;\; n>m\\
0\;\;\;\;n\le m
\end{gathered}
\right.
\end{align*}
There is the following relation between $S_m$ and $P_{l}^{\pm}$;
\begin{align}
S_mP_l^\pm=P_{m+l}^{\pm}S_m\quad
\label{spps}
\end{align}
The velocity projection ${\hat P}_v^-,{\hat P}_v^+$ are
defined in the Fourier representation as
\begin{align*}
[{\widehat {{\hat P}_v^-f}}](k)&\equiv
\left\{
\begin{gathered}
0\;\;\; k\in I_v\\
{\hat f}(k)\;\;\; k\in I_v^c,
\end{gathered}
\right.\\
[{\widehat{{\hat P}_v^+f}}](k)&\equiv
\left\{
\begin{gathered}
{\hat f}(k)\;\;\; k\in I_v\\
0\;\;\;\;k\in I_v^c
\end{gathered}
\right.
\end{align*}
where
\begin{align*}
I_v\equiv\left\{
k\in(-\pi,\pi];v<\sin k \right\},
\end{align*}\\
and $I_v^c$ is the complement of $I_v$.
By the definition of $T_t$ (\ref{tt}), we have
for $\vert n\vert < t(1-\delta)$,
\begin{align}
(T_t{\hat P}_v^{\pm}f)(n)
=(P_{-vt}^{\mp}T_tf)(n).
\label{tp}
\end{align}

Next we calculate the asymptotic value of  
\begin{align*}
\alpha_t(a^{\dag}(S_{vt}f))
=a^{\dag}({\rm e}^{ith}S_{vt}f).
\end{align*}
We claim that
\begin{align*}
\lim_{t\to\infty}
\Vert P_0^-S_{vt}{\rm e}^{ith}f-{\hat P}_v^+S_{vt}{\rm e}^{ith}f\Vert
=0
\end{align*}
We have to calculate
\begin{align*}
({\rm e}^{ith} S_{vt}f)(n)
=\frac{1}{2\pi}
\int_{-\pi}^{\pi}
dk {\rm e}^{-it(\cos k-\gamma)}{\rm e}^{i(n-vt)k}{\hat f}(k),
\end{align*}
which we obtain by replacing $n$ of (\ref{ho}), with $n-vt$.
Note that $S_m$ commutes with the dynamics ${\rm e}^{ith}$. 
Let us define $[g]_t$ as
\begin{align*}
\left[ g \right]_t(n)\equiv
\left\{
\begin{gathered}
(S_{vt}T_tg)(n)\quad \vert n-vt \vert \le t(1-\delta)\\
(S_{vt}{\rm e}^{ith}g)(n)\quad \vert n-vt \vert > t(1-\delta)
\end{gathered}
\right.
\end{align*}
Using the relation  (\ref{spps}), (\ref{tp}), we have
for $\vert n-vt \vert \le t(1-\delta)$,
\begin{align}
&\left[ {\hat P}_v^+g \right]_t(n)
=(S_{vt}T_t{\hat P}_v^+g)(n)\\
&=(S_{vt}P_{-vt}^-T_tg)(n)
=(P_0^-S_{vt}T_tg)(n)
=(P_0^-\left[g \right]_t)(n).
\label{pcom}
\end{align}
Let us fix some $\delta>0$.
Then, in the asymptotic limit $t\to\infty$,
we have from (\ref{pcom}),
\begin{align}
&\Vert P_0^-[g]_t-[P_v^+g]_t \Vert^2\nonumber\\
&=\sum_{\vert n-vt\vert > t(1-\delta)}
\vert (P_0^-S_{vt}{\rm e}^{ith}g(n)-
S_{vt}{\rm e}^{ith}{\hat P}_v^+g(n)\vert^2\nonumber\\
&+\sum_{\vert n-vt\vert\le t(1-\delta)}
\vert 
P_0^-[g]_t(n)-[{\hat P}_v^+g]_t(n)
\vert^2\nonumber\\
&\le 2\left[
\sum_{\vert n\vert> t(1-\delta)}
\vert {\rm e}^{ith}g(n)\vert^2
+\sum_{\vert n\vert> t(1-\delta)}
\vert {\rm e}^{ith}{\hat P}_v^+g(n)\vert^2
\right]\nonumber\\
&\to 2\left(B_{\delta}(g)+B_{\delta}\left({\hat P}_v^+g\right)\right).
\label{as1}
\end{align}
We also have in $t\to\infty$ limit,
\begin{align}
&\Vert S_{vt}{\rm e}^{ith}g-\left[g \right]_t\Vert^2\nonumber\\
&=\sum_{\vert n-vt\vert\le t(1-\delta)}
\vert (S_{vt}{\rm e}^{ith}g)(n)-(S_{vt}T_tg)(n)\vert^2\nonumber\\
&=\sum_{\vert n-vt\vert\le t(1-\delta)}
\vert ({\rm e}^{ith}g)(n-vt)-(T_tg)(n-vt)\vert^2\nonumber\\
&=\sum_{\vert n\vert\le t(1-\delta)}
\vert ({\rm e}^{ith}g)(n)-(T_tg)(n)\vert^2\nonumber\\
&=\sum_{\vert n\vert\le t(1-\delta)}
\vert (Ag)(t,n)\vert^2
\le \sum_{\vert n\vert\le t(1-\delta)}
\frac{(C^{\delta})^2}{t^2}\to 0
\label{as2}
\end{align}
Hence, we have 
\begin{align*}
&\lim_{t\to\infty}
\Vert P_0^-S_{vt}{\rm e}^{ith}g-{\hat P}_v^+S_{vt}{\rm e}^{ith}g \Vert\\
&\le \lim_{t\to\infty}
\Vert P_0^-S_{vt}{\rm e}^{ith}g-P_0^-[g]_t\Vert\\
&+\lim_{t\to\infty}\Vert P_0^-[g]_t-[P_v^+g]_t\Vert
+\lim_{t\to\infty}\Vert [P_v^+ g]_t -{\hat P}_v^+S_{vt}{\rm e}^{ith}g \Vert\\
&\le 2(B_{\delta}(g)+B_{\delta}({\hat P}_v^+g)).
\end{align*}
We used (\ref{as1}), (\ref{as2}) and the commutativity
of ${\hat P}_v^+$, $S_{vt}$, and ${\rm e}^{ith}$.
Note that the first term $\lim_{t\to\infty}
\Vert P_0^-S_{vt}{\rm e}^{ith}g-{\hat P}_v^+S_{vt}{\rm e}^{ith}g \Vert$
is $\delta$-independent.
So, taking $\delta\to 0$, we have
\begin{align}
\lim_{t\to\infty}
\Vert P_0^-S_{vt}{\rm e}^{ith}g-{\hat P}_v^+S_{vt}{\rm e}^{ith}g \Vert=0.
\label{asym1}
\end{align}
Similarly, we have
\begin{align}
\lim_{t\to\infty}
\Vert P_0^+S_{vt}{\rm e}^{ith}g-{\hat P}_v^-S_{vt}{\rm e}^{ith}g \Vert=0.
\label{asym2}
\end{align}

Next, we substitute the above result to the initial state to
derive the two point function $\omega_v(a^{\dag}(g)a(f))$.
For the purpose, we rewrite the initial state
with the projection operators $P_{\pm}$.
Note that
\begin{align*}
{\tilde f}_{-}(k)&=-i\sum_{n\le 0}f_n\sin(n-1)k\\
&=-i\sum_{n\le 0}f_n\frac{{\rm e}^{i(n-1)k}-{\rm e}^{-i(n-1)k}}{2i}\\
&=-\frac{1}{2}\left[ {\widehat{P_0^-f}}(-k){\rm e}^{-ik}
-{\widehat{P_0^-f}}(k){\rm e}^{ik}\right]
\end{align*}
Hence we have
\begin{align}
&\omega_-(a^{\dag}(g)a(f))\\
&=\frac{1}{4\pi}\int_{-\pi}^{\pi}
dk \left[
{\overline{\widehat{P_0^-f}}}(-k){\widehat{P_0^-g}}(-k)
+{\overline{\widehat{P_0^-f}}}(k){\widehat{P_0^-g}}(k)
\right.\nonumber\\
&\left.
-{\rm e}^{2ik}{\overline{\widehat{P_0^-f}}}(-k){\widehat{P_0^-g}}(k)
-{\rm e}^{-2ik}{\overline{\widehat{P_0^-f}}}(k){\widehat{P_0^-g}}(-k)
\right]\rho_-(k)
\label{rewrite}
\end{align}
We are now interested in 
$\omega_-(a^{\dag}(g_t)a(f_t))$,
with $g_t=S_{vt}{\rm e}^{ith}g$, $f_t=S_{vt}{\rm e}^{ith}f$.
We can see from (\ref{rewrite}), that $\omega_-(a^{\dag}(g_t)a(f_t))$ is written with $P_0^-S_{vt}{\rm e}^{ith}g$ and 
$P_0^-S_{vt}{\rm e}^{ith}f$.
Hence, we can apply the asymptotic form (\ref{asym1}).\\
Applying the formula, 
we have
\begin{align*}
&\lim_{t\to\infty}
\omega_-(a^{\dag}(g_t)a(f_t))\\
&=\lim_{t\to\infty}
\frac{1}{4\pi}\int_{-\pi}^{\pi}
dk \left[2{\hat P}_v^+(k){\bar{\hat{f}}}(k){\hat{g}}(k)\right.\\
&-{\rm e}^{-2ik}{\hat P}_v^+(k){\hat P}_v^+(-k){\rm e}^{2ivtk}
{\bar{\hat{f}}}(k){\hat{g}}(-k)\\
&\left.
-{\rm e}^{2ik}{\hat P}_v^+(k){\hat P}_v^+(-k){\rm e}^{-2ivtk}
{\bar{\hat{f}}}(-k){\hat{g}}(k)
\right]\rho_-(k).
\end{align*}
Here we used the fact that the Fourier representation of $S_m$ is
\begin{align*}
{\widehat {S_mf}}(k)={\rm e}^{-imk}{\hat f}(k).
\end{align*}
By the Riemann-Lebesgue Theorem,
the second and the third term vanishes and
we finally obtain
\begin{align*}
\lim_{t\to\infty}\omega_-(a^{\dag}(g_t)a(f_t))
=
\frac{1}{2\pi}\int_{v<\sin k}
dk \rho_-(k){\bar{\hat{f}}}(k){\hat{g}}(k)
\end{align*}
Similarly, we obtain
\begin{align*}
\lim_{t\to\infty}\omega_+(a^{\dag}(g_t)a(f_t))
=
\frac{1}{2\pi}\int_{v\ge\sin k}
dk \rho_+(k){\bar{\hat{f}}}(k){\hat{g}}(k)
\end{align*}
Hence, the explicit form of the two point function of 
$\omega_v(a^{\dag}(g)a(f))$ is
\begin{align*}
\omega_v(a^{\dag}(g)a(f))
=\frac{1}{2\pi} 
 \int_{v<\sin k} dk
\rho_-(k){\bar{\hat{f}}}(k){\hat{g}}(k)\\
+\frac{1}{2\pi}\int_{v\ge\sin k} dk
\rho_+(k){\bar{\hat{f}}}(k){\hat{g}}(k).
\end{align*}
Note that $\omega_v(a^{\dag}(g)a(f))$ is translation invariant.

The velocity of the particle with momentum $k$ is $\sin k$.
We see that only the particles at temperature $1/\beta_-$
with velocity $(v\le \sin k \le 1)$ contribute to 
the state $\omega_v$.
This represents the situation that on the inertial system 
which moves with velocity $v$, 
quasi particles with velocity less than $v$
in the left part of the chain go to left infinity
and do not appear in the correlation function $\omega_v$.
This feature is also the case for the particles with temperature
$1/\beta_+$
\section{The asymptotic profile}\label{profile}
With the two-point function, we can calculate 
the asymptotic profile of physical quantities.
The magnetization profile 
$m(v)$ is 
\begin{align*}
m(v)=\frac{1}{2\pi} 
\int_{v>\sin k} dk
\rho_+(k)\\
+\frac{1}{2\pi} \int_{v\le\sin k} dk
\rho_-(k)-\frac{1}{2}.
\end{align*}
The magnetic current profile 
is defined by 
\begin{align*}
J^M(v)=\lim_{t\to\infty}\omega(\alpha_t(J_{vt}^M)),
\end{align*} 
where $J_n^M=S_n^yS_{n+1}^x-S_n^xS_{n+1}^y$ is the magnetic current 
at the site $n$.
Similarly, $J^M(v)$ is calculated as 
\begin{align*}
J^M(v)=\frac{1}{2\pi} 
\int_{v>\sin k} dk
\rho_+(k)\sin k\\
+\frac{1}{2\pi} \int_{v\le\sin k} dk
\rho_-(k)\sin k
\end{align*}

Let us consider the zero-temperature case:
each side is in the ground state ($\beta_+=\beta_-=\infty$)
with magnetic field $\gamma_+$,
$\gamma_-$ respectively.
We further assume $\gamma\equiv\gamma_+=-\gamma_-$.
This is the situation that was considered in \cite{ant2},
and we confirm their results,
\begin{align*}
m(v)=-m(-v)
=\left\{
\begin{gathered}
0\;\;\; \;\;\; \;\;\; \;\;\;\;\;\; \;\;\; \;\;\;\;\;\; \;\;\; \;\;\; 0\le v<\cos\pi m_0\\
-m_0+\frac{\arccos(v)}{\pi}\;\;\; \cos\pi m_0\le v<1\\
\frac12 -m_0\;\;\; \;\;\; \;\;\;\;\;\; \;\;\; \;\;\;\;\;\; \;\;\; \;\;\; \;\;\;\;\;\; 1\le v
\end{gathered}
\right.
\end{align*}
where $\gamma=\sin\pi m_0$.
\begin{align*}
J^M(v)=J^M(-v)
=\left\{
\begin{gathered}
\frac{1}{\pi}\gamma=\frac{1}{\pi}\sin\pi m_0
\;\;\;\;\; 0\le v<\cos\pi m_0\\
\frac{1}{\pi} \sqrt{1-v^2}\;\;\; \cos\pi m_0\le v <1\\
0\;\;\; \;\;\; \;\;\;\;\;\; \;\;\; \;\;\;\;\;\; 1\le v
\end{gathered}
\right.
\end{align*}
In the zero-temperature case,
regardless of the strength of the external fields $\gamma_+$, $\gamma_-$,
we can show that the magnetization profile $m(v)$
is monotone. 
The situation is classified by the following conditions:
(A) the absolute value of the external field:
(i)$\vert \gamma_-\vert,\vert \gamma_+\vert \le 1$,
(ii)$\vert \gamma_-\vert\le 1,\vert \gamma_+\vert >1$
or $\vert \gamma_+\vert\le 1,\vert \gamma_-\vert >1$,
(iii)$\vert \gamma_-\vert ,\vert \gamma_+\vert >1$
and
(B) the signs of the external field:
(i)${\rm sign}(\gamma_+)={\rm sign}(\gamma_-)$,
(ii)${\rm sign}(\gamma_+)=-{\rm sign}(\gamma_-)$.
Hence, we have $3\times 2=6$ cases.
By the explicit calculation, we can show that for all the situation,
the magnetization profile $m(v)$ is monotone. 
Fig.3 shows the magnetization profile for each case. 
As seen in the following, the finiteness of the temperature
destroies this monotonicity when the external field is small.

\begin{figure}[b]
\begin{center}
\includegraphics[height=2cm,clip]{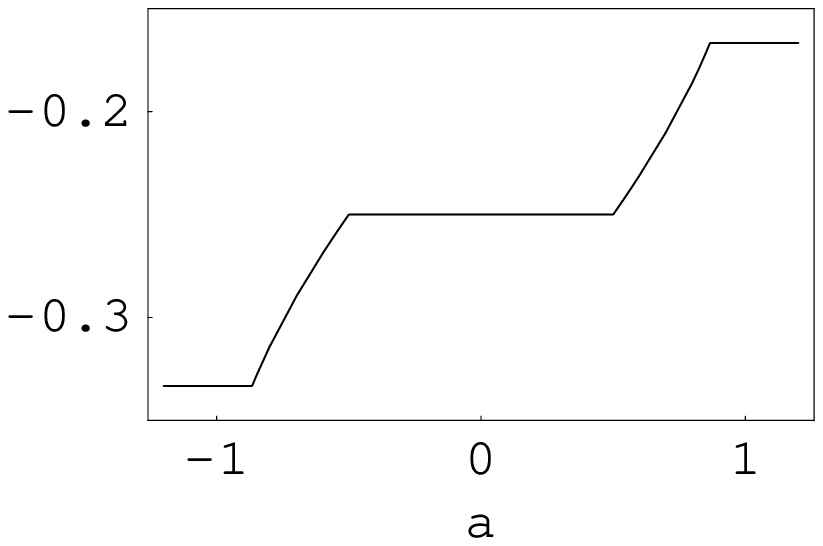}
\includegraphics[height=2cm,clip]{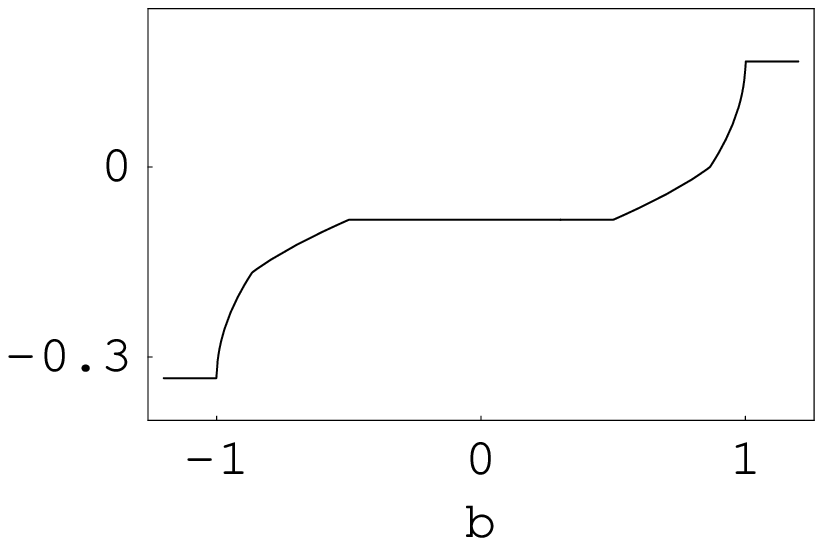}
\includegraphics[height=2cm,clip]{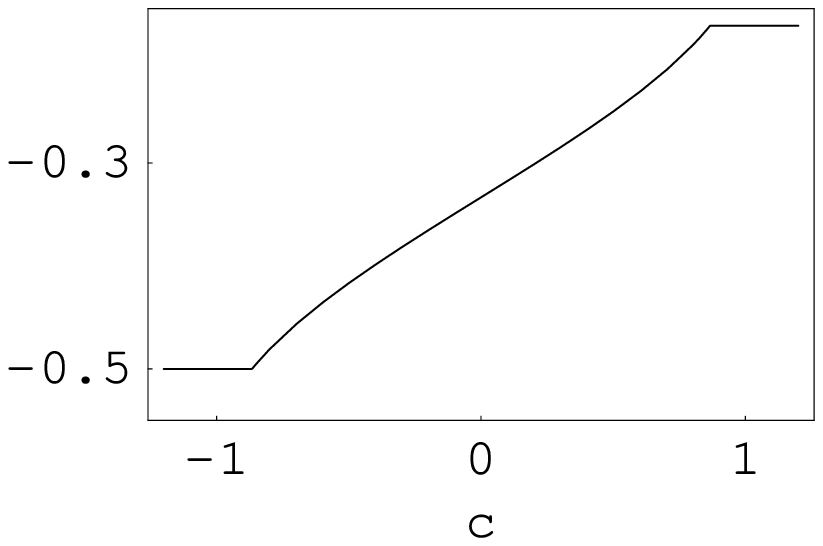}
\includegraphics[height=2cm,clip]{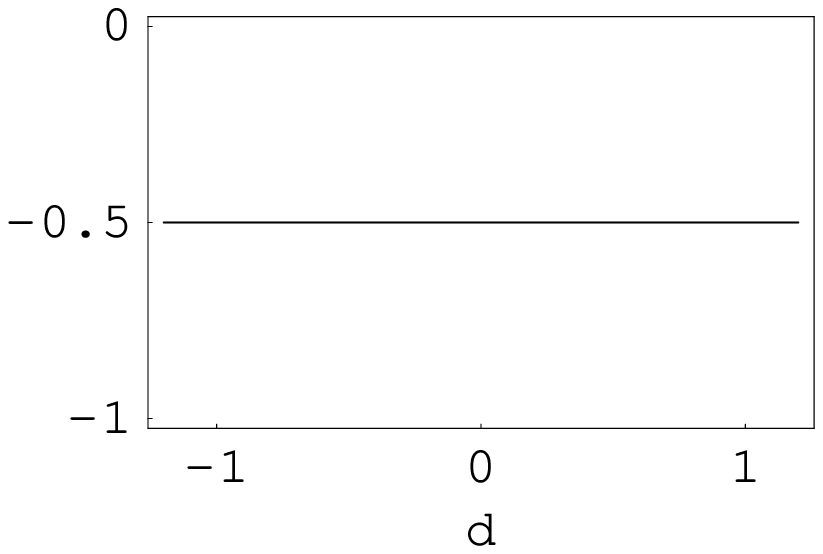}
\includegraphics[height=2cm,clip]{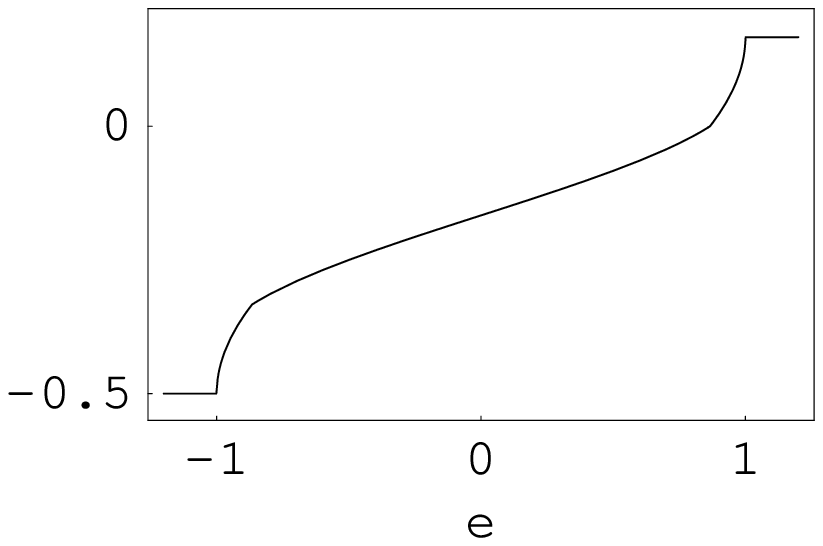}
\includegraphics[height=2cm,clip]{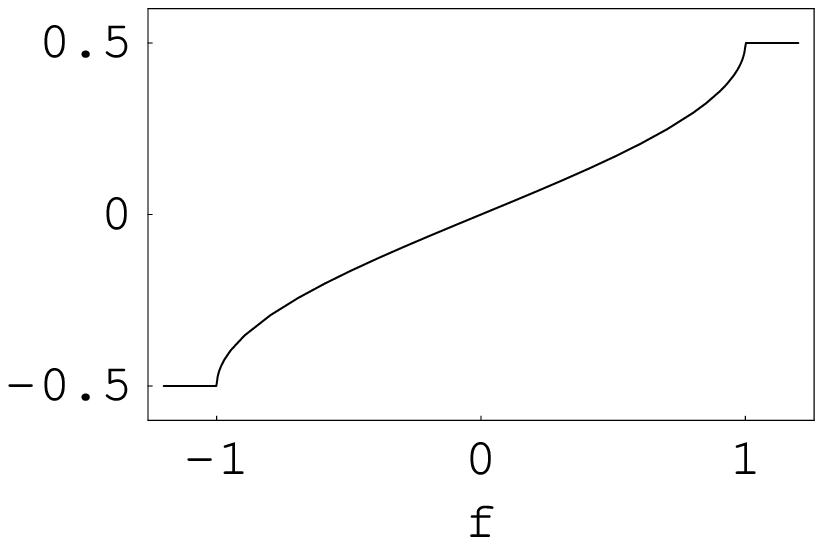}
\begin{picture}(1,0)
 \put(-215,-5){{
 Fig.3:The magnetization profile at zero temperature 
 for various}}
\put(-215,-15){{
 external fields.
The $x$-axes is the scaling factor $v$.
The $y$-axes is}}
 \put(-215,-25){{  the value of magnetization.
 All the case is classified by (A)$\times$(B):}}
\put(-215,-35){{ 
(A) the 
absolute value of the external field:
(i)$\vert \gamma_-\vert,\vert \gamma_+\vert \le 1$,
}}
\put(-215,-45){{
(ii)$\vert \gamma_-\vert\le 1$,$\vert \gamma_+\vert >1$
or $\vert \gamma_+\vert\le 1,\vert \gamma_-\vert >1$,
(iii)$\vert \gamma_-\vert ,\vert \gamma_+\vert >1$,
}}
\put(-215,-55){{
(B) the signs of 
the external field:
(i)${\rm sign}(\gamma_+)={\rm sign}(\gamma_-)$,
(ii)}}
\put(-215,-65){{
${\rm sign}(\gamma_+)=-{\rm sign}(\gamma_-)$.
 (a):A-(i),B-(i),
 (b):A-(i),B-(ii),}}
\put(-215,-75){{
(c):A-(ii),B-(i),
(d):A-(iii),B-(i),
(e):A-(ii),B-(ii),(f):A-(iii),B-(ii).
}}
 \end{picture}
\end{center}
\end{figure}
\begin{figure}
\begin{center}
\includegraphics[height=2cm,clip]{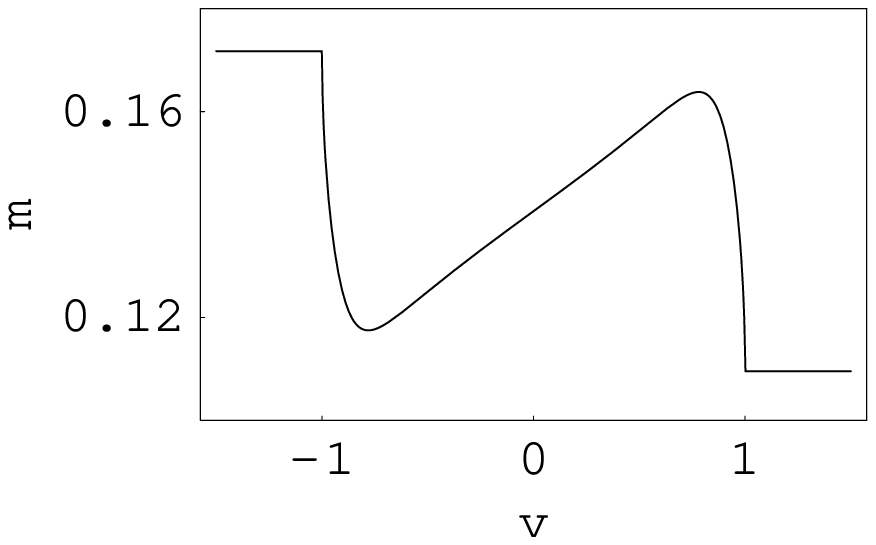}
\includegraphics[height=2cm,clip]{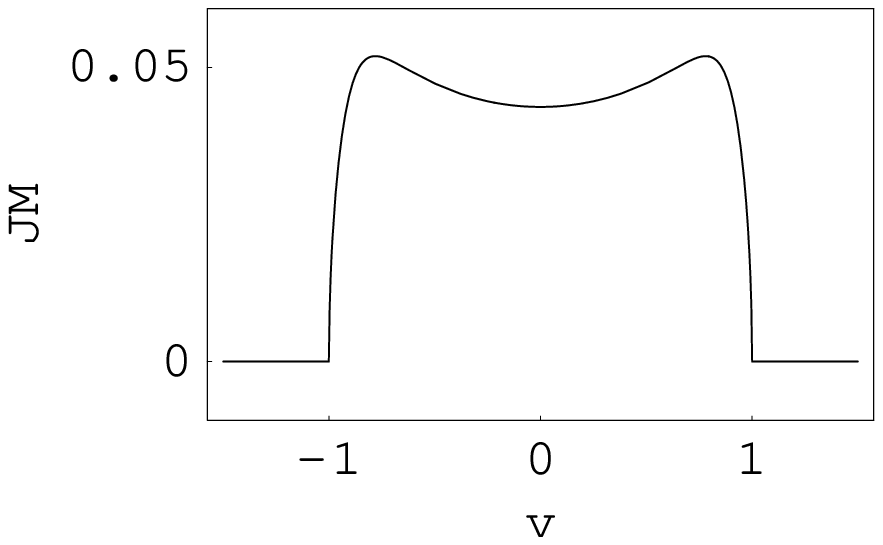}
\includegraphics[height=2cm,clip]{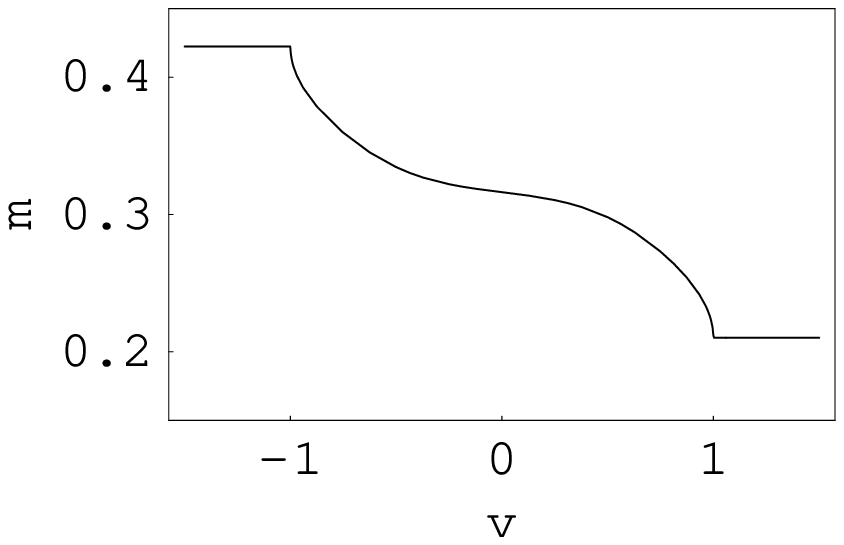}
\includegraphics[height=2cm,clip]{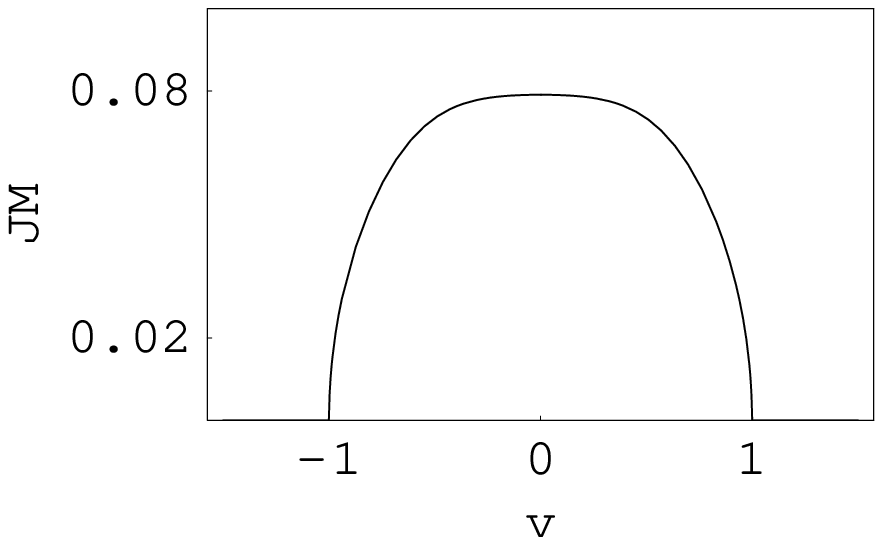}
\includegraphics[height=2cm,clip]{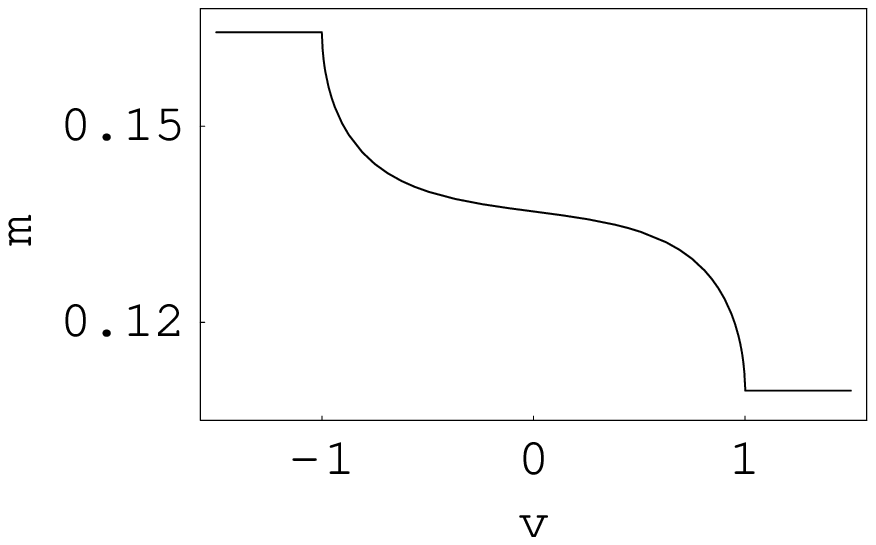}
\includegraphics[height=2cm,clip]{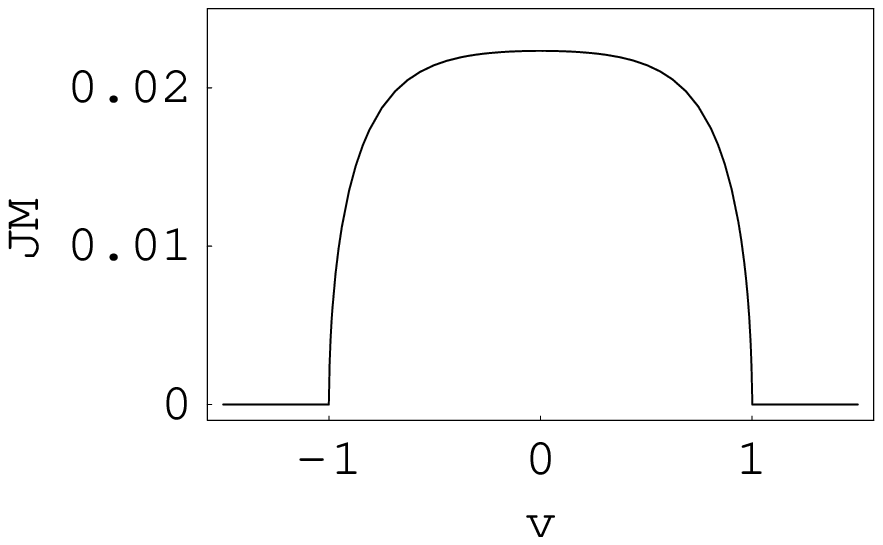}
\begin{picture}(500,65)(-220,-60)
\put(-205,152){{\tiny (a)}}
\put(-205,95){{\tiny (b)}}
\put(-205,35){{\tiny (c)}}
\put(-215,-10){{\small
 Fig.4:The profile of the magnetization and the magnetic current }}
\put(-215,-20){{\small at finite temperature for various
 external fields.
 The left graph }}
\put(-215,-30){{\small
shows the magnetization profile, and the right one shows the 
}}
\put(-215,-40){{\small magnetic current profle.
In the left graph,
$x$-axes is the scaling}}
\put(-215,-50){{\small
factor $v$
and the $y$-axes
is the value of the magnetization $m(v)$. }}
\put(-215,-60){{\small
In the
right graph,
$x$-axes is the scaling factor $v$ and the $y$-axes
}}
\put(-215,-70){{\small is the
 value of the magnetic current $J^M(v)$. 
(a):$\gamma=-0.5,\beta_-=10$,}}
\put(-215,-80){{\small $\beta_+=1$,
(b):$\gamma=-1,\beta_-=10,\beta_+=1$,
(c):$\gamma=-0.5,\beta_-=2,\beta_+=1$.}}

\end{picture}
\end{center}
\end{figure}

To concentrate on the thermal inhomogeneity,
let us consider the situation $\gamma\equiv\gamma_+=\gamma_-\ne 0$.
Due to the non-zero $\gamma$, the spins have finite magnetization
up to the value of the temperature.
The left side of Fig.4 shows the profile 
of the magnetization for various 
values of the external field and temperature.
Fig.(a) corresponds to the case $\gamma=-0.5,\beta_-=10,\beta_+=1$.
It shows the non-monotone profile.
As the magnetization profile is monotone 
in zero temperature $\beta_-=\beta_+=\infty$,
this is considered to be a purely thermal property.
When we increase the strength of $\gamma$,
the non-monotonicity is lost.
Fig.(b) shows the $\gamma=-1$ case with the same temperature:
$\beta_-=10,\beta_+=1$.
We can see the monotone profile.
On the other hand, decrease of the difference of the
temperature also destroies the non-monotonicity.
Fig.(c) show the $\gamma=-0.5$ case with the different temperature:
$\beta_-=2,\beta_+=1$.
The difference emerges also in the magnetic current.
The right side of Fig.4 shows 
the profile of the magnetic current.
In the case (a), the current takes the maximum value at
two points.
They corresponds to the two extremum point of the magnetization profiles.
On the other hand, in the monotone case,
the current takes the maximum at the origin $v=0$.

The property of the profile (monotone/non-monotone)
can be explained by the velocity distribution.
Firstly, recall that the velocity of the particle
at momentum $k$ is $\sin k$.
To see the dependence of the velocity distribution, 
let us consider the derivatives of $m(v)$ and
$J^M(v)$ with $v$.
For simplicity, we restrict ourselves to the case
that both of $\rho_-(k)$, $\rho_+(k)$ are continuous,
i.e, the both sides are initially in finite temperature.
In this case, we can differentiate $m(v)$ and  $J^M(v)$ with $v$
for $-1<v<1$.
We have
\begin{align*}
\frac{dm(v)}{dv}=
\frac{1}{\sqrt{1-v^2}}\left[
p_+(v)-p_-(v)
\right],
\end{align*}
\begin{align*}
\frac{dJ^M(v)}{dv}=
\frac{v}{\sqrt{1-v^2}}\left[
p_+(v)-p_-(v)
\right],
\end{align*}
where $p_-$(resp.$p_+$) is the
velocity distribution,
\begin{align*}
p_r(v)\equiv
\frac{1}{2\pi}\left[
\frac{1}{1+e^{-\beta_{r}
\left( \sqrt{1-v^2}-\gamma \right)}}
+\frac{1}{1+e^{-\beta_{r}
\left(-\sqrt{1-v^2}-\gamma \right)}}\right]
\end{align*}
From these expressions, we can see that the difference
$p_+(v)-p_-(v)$ determines the monotone/non-monotone
of $m(v)$ and $j(v)$;
if, for example, there exists a point where
$p_+(v)>p_-(v)$ changes to $p_+(v)<p_-(v)$,
the profile is non-monotone.
Fig.5 shows $p_-(v)$, $p_+(v)$  
for the above mentioned situation.
We can see the crossing only for the case of non-monotone (a).
\section{Discussion}
We have investigated the profiles of the magnetization and
the magnetic current, in the intermediate time 
towards the non-equilibrium steady state,
using the transversed $XX$-model.
We have found an interesting property:
depending on the strength of the external fields
and the values of initial temperature,
the profile shows monotone/non-monotone property.
This emerges as the result of the initial velocity distribution 
of the right and the left side.
If there is a crossing between two distributions,
the profile becomes non-monotone.
This initial velocity dependence is due to a fact that 
the transverse $XX$-model
preserves the one-particle mode.
Each particle runs to the infinity with its own velocity.
In this sense, the integrability affects the diffusion profile
in an essential way.

The derivation of the asymptotic profile is
carried out by showing the equation (\ref{asym1}),(\ref{asym2}).
These equations are due to (\ref{tp});
the fact that for each site,
there is only the particle with specific momentum.
The specific momentum is the momentum where
the phase velocity is stational
(\ref{dash}).
In other words, if the dynamics of free Fermion is asymptotically
dominated by the stationary point,
we would have the same property as in this paper,
even if the dispersion is not cosine. 
\acknowledgements{The author would like to thank Prof.M.Wadati for valuable comments and critical reading of the manuscript,
and also thank Dr.K.Saito for helpful discussions.}
\begin{figure}[b]
\begin{center}
\includegraphics[height=2cm,clip]{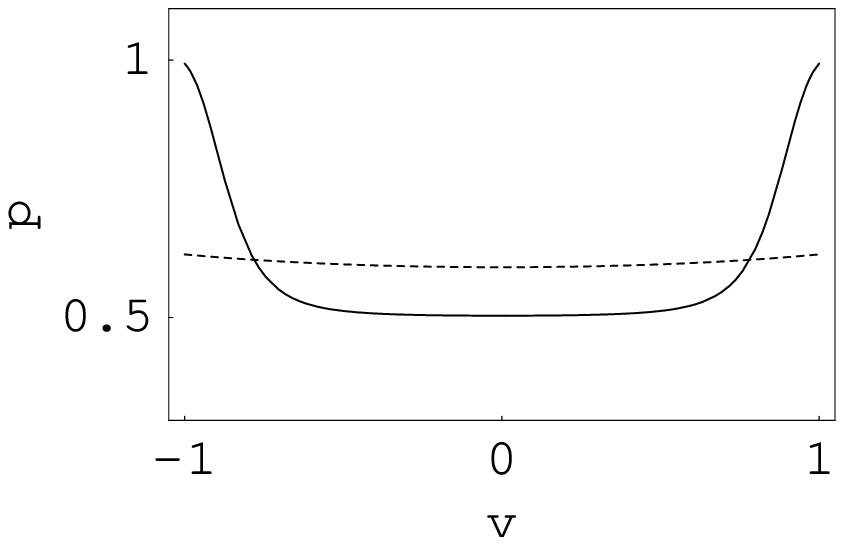}
\includegraphics[height=2cm,clip]{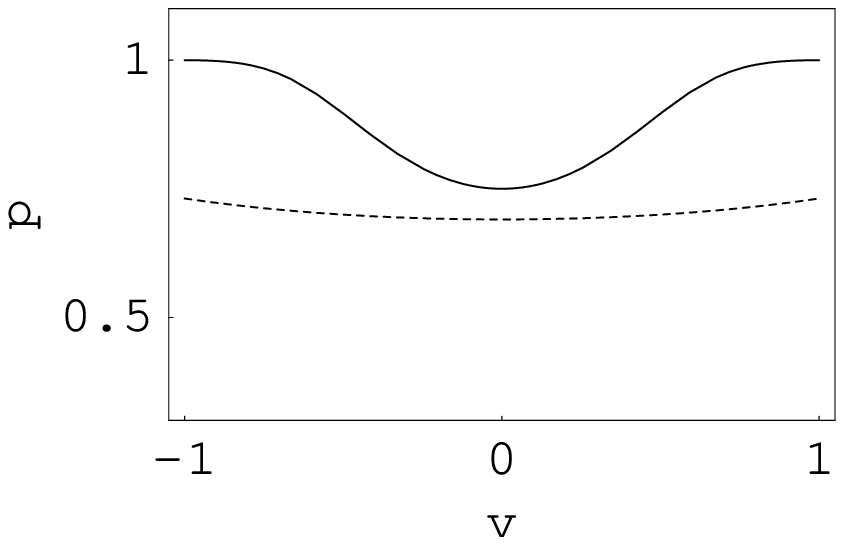}
\includegraphics[height=2cm,clip]{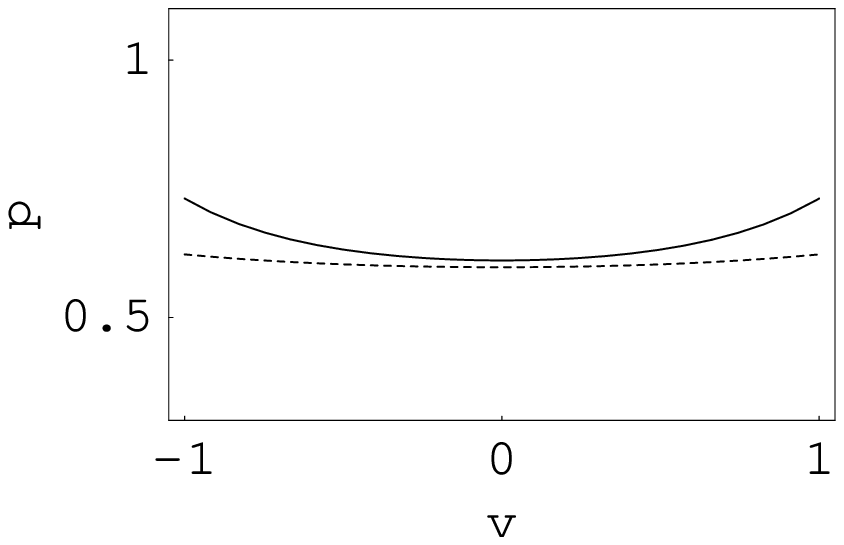}
\begin{picture}(500,65)(-220,-60)
\put(-205,95){{\tiny (a)}}
\put(-103,95){{\tiny (b)}}
\put(-150,35){{\tiny (c)}}
\put(-215,-20){{\small
 Fig.5:The velocity distribution of the right and left sides.
The}}
\put(-215,-30){{\small  
 $x$-axes is the velocity $v$,
the $y$-axes is the distributuion $p(v)$. 
}}
\put(-215,-40){{\small 
The solid line shows $p_-(v)$,
and the dash line, $p_+(v)$.
}}
\put(-215,-50){{\small (a):$\gamma=-0.5$,$\beta_-=10,\beta_+=1$,
(b):$\gamma=-1,\beta_-=10,\beta_+=1$,}}
\put(-215,-60){{
(c):$\gamma=-0.5$, $\beta_-=2,\beta_+=1$.}}
\end{picture}
\end{center}
\end{figure}

\end{document}